\documentstyle[12pt]{article}
\begin{document}
\[\]
\begin{center}
{\large \bf Spherically Symmetric Thick Branes in Vacuum}
\[\]
{\bf R. Mansouri\footnote{E-mail address: mansouri@sharif.edu}, M.
Borhani\footnote{E-mail address: borhani@physics.sharif.edu}, S.
Khakshournia\footnote{E-mail address:
skhakshour@seai.neda.net.ir}}\\
Department of Physics, Sharif University of Technology,\\
P.O.Box 11365--9161, Tehran, Iran\\
Institute for Studies in Theoretical Physics and Mathematics,\\
P.O.Box 19395--5531, Tehran, Iran
\end{center}
\[\]
\begin{center}
{\bf \ Abstract}
\end{center}
\hspace*{0.5cm}We consider a spherical thick 3-brane immersed in a
five-dimensional bulk spacetime. We demonstrate how the thick
brane equation of motion expanded in powers of the thickness of
the brane can be obtained from the expected junction conditions on
the boundaries of thick brane with the two embedding spacetimes.
It is shown that the finite thickness leads to a faster collapse
of the spherical shell.
\newpage
\subsection*{I. Introduction}
\hspace*{0.5cm}There has been much recent interest in the idea
that our universe may be a 3-brane(a hypersurface) embedded in
some higher dimensional spacetime. The standard model fields are
restricted to our 3-brane whereas gravity propagates in the bulk.
Such models were suggested to explain the large gap between the
scale of gravity $M_{Pl}=(8\pi G_{N})^{-1/2}$ and the scale of the
other interactions so that the "true" Planck scale can  be
diminished to the TeV scale by introducing extra dimensions(see
[1] for a recent review). Usually the brane is considered to be
infinitely thin along this extra-dimension. This simplification
has the disadvantage that the curvature is singular at the
location of the brane. In fact, thin brane approximation is valid
as long as the energy scales of the model are much smaller than
the energy scale related to the inverse thickness of the brane so
that we may not neglect the thickness of the brane at the
fundamental scale. Authors usually treat a thick brane as an
object made of a real scalar field and investigate its dynamics by
solving the coupled Einstein-scalar equations  with a suitable
potential for the scalar field in the context of the classical
field theory[2-4](see also [5] for a recent study
in 4D gravity).\\
 In this paper we consider an inhomogeneous spherically symmetric
thick 3-brane separating two metrically  different
five-dimensional regions of the bulk assumed to be empty. The
matter within the brane is assumed to be a pressure-free ideal
fluid. By writing down the proper junction conditions on the
boundaries of the brane with the bulk, we obtain an equation of
motion for the brane to determine the effect of the thickness on
its motion without having to calculate the detailed dynamics of
its internal structure. Recently, a similar treatment was worked
out for the case of a spherical thick shell in vacuum[6].\\
\subsection*{II. Thick Brane Model}
\hspace*{0.5cm}Consider a spherical thick brane with two
boundaries $\Sigma_{1}$ and $\Sigma_{2}$ dividing the five
dimensional spacetime into three-regions: ${\cal M}_{in}$ for
inside the inner boundary $\Sigma_{1}$,  ${\cal M}_{out}$ for
outside the outer boundary $\Sigma_{2}$, and ${\cal M}$  for the
thick brane having two boundaries $\Sigma_{1}$ and $\Sigma_{2}$.
We start by choosing the five dimensional Lemaitre-Tolman-Bondi
(LTB) metric to describe the brane-universe as a spherically
symmetric thick shell of the dust, written here in the
synchronized comoving coordinates $(\tau,r,\chi,\theta,\phi)$
(see [7] for 4D gravity)
\begin{equation}
\label{metric} ds^2\Bigl|_{{\cal M}}=-d\tau^2+
\frac{R^{'2}}{1+E(r)}dr^2+R^2(r,\tau) \left(d\chi^2+sin^2\chi
(d\theta^2+ sin^2\theta d\phi^2)\right),
\end{equation}
where overdot and prime denote partial differentiation with
respect to $\tau$ and $r$, respectively, and $E(r)$ is an
arbitrary real function of integration such that $E(r)>-1 $. The
five dimensional Einstein field equations take the usual form
\begin{equation}
\check{G}_{\mu\nu}= -8\pi G_{5}\check{T}_{\mu\nu}.
\end{equation}
To be compatible with the spacetime symmetries (1), the five
dimensional energy-momentum tensor of the dust matter content in
the thick brane must be necessarily of the form
\begin{equation}
\check{T}^{\mu}_{\nu}=diag(-\check{\rho},0,0,0,0),
\end{equation}
where $\check{\rho}(r,\tau)$ is the five dimensional energy
density of the matter fluid in ${\cal M}$. From the metric (1),
the non-vanishing components of the Einstein tensor
$\check{G}_{\mu\nu}$ are computed as follows
\begin{equation}
\check{G}_{00}=\frac{3\left(-2R\dot{R}\dot{R}'-2R'\dot{R}^2+RE'+2R'E(r)\right)}{2R^2R'},
\end{equation}
\begin{equation}
\check{G}_{11}=\frac{R'^2\left(3R\ddot{R}+3\dot{R}^2-3E(r)\right)}{R^2\left(1+E(r)\right)},
\end{equation}
\begin{equation}
\check{G}_{22}=R'^{-1}\left(2R\ddot{R}R'+2R\dot{R}\dot{R'}-RE'(r)+R'\dot{R}^2-R'E(r)+R^2\ddot{R'}\right),
\end{equation}
\begin{equation}
\check{G}_{33}=sin^{2}\chi\check{G}_{22},
\end{equation}
\begin{equation}
\check{G}_{44}=\sin^{2}\theta\check{G}_{33}.
\end{equation}
On substituting into the Einstein equations (2) supplemented by
the Bianchi identity ${T^{\mu}}_{\nu;\mu}=0$, and after a little
algebra, we end up with the following two independent field
equations:
\begin{equation}
\label{dynamics1} \dot{R}^2(r,\tau)=E(r)+\frac{F(r)}{R^2},
\end{equation}
\begin{equation}
\label{dynamics2}16\pi G_{5}
\check{\rho}(r,\tau)=\frac{3F^{'}(r)}{R^3 R^{'}},
\end{equation}
where $F(r)$ is another arbitrary real smooth function of
integration such that $ F(r)>0$. Furthermore, in order to avoid
shell crossing of dust matter during their radial motion, we
require $R^{'}(r,t)>0 $. This together with an assumption of
positive energy density $\rho(r,\tau)>0 $, implies that
$F^{'}(r)\geq0 $. The induced intrinsic metric of $
\Sigma_{j}(j=1,2) $ may be written as
\begin{equation}
\label{boundary} ds^2\mid_{\Sigma_{j}}=-d\tau^2+R_{j}^2(\tau_{j})
\left(d\chi^2+sin^2 \chi(d \theta^2+ sin^2 \theta d
\phi^2)\right),
\end{equation}
where $R_{j}(\tau_{j})$ being the proper radius of $ \Sigma_{j}
$. For simplicity, we may assume that the boundaries $\Sigma_{1} $
and $ \Sigma_{2} $ are comoving with respect to the LTB geometry.
This requires that the peculiar velocity of $ \Sigma_{j}$
measured by the comoving observers of ${\cal M}$ to be zero. Then
the matching relations yield
\begin{equation}
\label{matching} \tau=\tau_{j}+const.
\end{equation}
Now, let us define the constant comoving thickness of the brane as
follows
\begin{equation}
\label{thickness} 2\delta =r_{2}-r_{1},
\end{equation}
where $r_{1}$ and $r_{2}$ are comoving radii of the boundaries $
\Sigma_{1}$ and $\Sigma_{2}$, respectively.\\
We naively expect the continuity of the second fundamental form of
$\Sigma_{j}$, or the extrinsic curvature tensor $K_{ab}$
 of $\Sigma_{j}$, so that $\Sigma_{1} (\Sigma_{2})$ would be a
 boundary surface separating ${\cal M}$ region from ${\cal M}_{in}$
 (${\cal M}_{out}$). This crucial requirement is formulated as [6]
\begin{equation}
\left[ K_{ab} \right] \stackrel{\Sigma_{j}}{=} 0
\quad\quad\quad\quad
 \quad \quad (j=1,2),
\end{equation}
where the square bracket indicates the jump of $K_{ab}$ across
$\Sigma_{j}$, Latin indices range over the intrinsic coordinates
of $\Sigma_{j}$ denoted by $(\tau_{j}, \chi, \theta , \varphi )$,
where $\tau_{j}$ is the proper time of $\Sigma_{j}$. In
particular, the angular component of Eq. (14) on each boundary is
written as
\begin{eqnarray} K_{\chi}^{\chi^{+}}
\Bigl|_{\Sigma_{1}} -K_{\chi}^{\chi^{-}}
\Bigr|_{\Sigma_{1}}=0 , \\
K_{\chi}^{\chi^{+}} \Bigl|_{\Sigma_{2}} -K_{\chi}^{\chi^{-}}
\Bigr|_{\Sigma_{2}}=0 ,
\end{eqnarray}
where the superscript $+(-)$ refers to the side of $\Sigma_{j}$
towards which the corresponding unit spacelike normal vector
$n^{\alpha} (-n^{\alpha})$ points. This means that on $\Sigma_{1}
(\Sigma_{2})$, the superscript + refers to the region ${\cal M}
({\cal M}_{out})$ and the superscript $-$ refers to the region
${\cal M}_{in}(\cal M )$. The addition of Eqs. (15) and (16),
yields the following equation
\begin{equation} K_{\chi}^{\chi^{+}}
\Bigl|_{\Sigma_{2}} -K_{\chi}^{\chi^{-}}
\Bigl|_{\Sigma_{1}}+K_{\chi}^{\chi^{+}} \Bigl|_{\Sigma_{1}}-
K_{\chi}^{\chi^{-}} \Bigl|_{\Sigma_{2}}=0.
\end{equation}
In our case, according to the five dimensional Birkhoff theorem [8
], the bulk space-time exterior to the brane is five dimensional
Schwarzschild, and the interior is taken to be five dimensional
Minkowski flat spacetime:
\begin{equation}
\label{metric} ds^2\Bigl|_{{\cal
M}_{out}}=-U(r)dt^2+U(r)^{-1}dr^2+r^2\left(d\chi^2+sin^2\chi
(d\theta^2+ sin^2\theta d\phi^2)\right),
\end{equation}
\begin{equation}
\label{metric} ds^2\Bigl|_{{\cal M}_{in}}=-dt^2+
dr^2+r^2\left(d\chi^2+sin^2\chi (d\theta^2+ sin^2\theta
d\phi^2)\right),
\end{equation}
with
\begin{equation}
U(r)\equiv1-\frac{{\cal R}^2(r)}{r^2}.
\end{equation}
Different terms appearing in the equation $(17)$ may now be
explicitly calculated. Given the metric $(1)$, field equations
$(9,10)$, and using Eq. $(12)$, we can obtain the relevant
extrinsic curvature tensors in the brane region ${\cal M}$ as
\begin{equation}
\label{ic1} K^{\chi+}_{\chi}\Bigl|_{\Sigma_{1}}=
\frac{1}{R_{1}}\left(1+\dot{R}_{1}^2-\frac{
F(r_{1})}{R_{1}^2}\right)^{\frac12} , \hspace{0.8cm}
K^{\chi-}_{\chi}\Bigl|_{\Sigma_{2}}=\frac{1}{R_{2}}
  \left(1+\dot{R}_{2}^2-\frac{F(r_{2})}{R_{2}^2}\right)^{\frac{1}{2}},
\end{equation}
where $ R_{j}=R(r_{j},t) $. Furthermore,  the relevant extrinsic
curvature tensors in $ {\cal M}_{in}$ and ${\cal M}_{out} $ are
also calculated as
\begin{equation}
\label{ic2}
K^{\chi-}_{\chi}\Bigl|_{\Sigma_{1}}=\frac{1}{R_{1}}(1+\dot{R}_{1}^2)^\frac{1}{2}
, \hspace{0.8cm}K^{\chi
+}_{\chi}\Bigl|_{\Sigma_{2}}=\frac{1}{R_{2}}
\left(1+\dot{R}_{2}^2-\frac{{{\cal
R}}^2(r_{2})}{R_{2}^2}\right)^{\frac{1}{2}},
\end{equation}
where ${\cal R}(r_{2})$ is the Schwarzschild radius of the five
dimensional spherical brane within the comoving surface $ r_{2}
$. Now, to obtain the dynamical equation of the thick brane, we
first expand the following quantities in a Taylor series around $
r_{0} $, the mean comoving radius of the thick brane
\begin{equation}
R(r_{j},\tau)=R(r_{0},\tau)+\epsilon_{j} \delta
R^{'}(r_{0},\tau)+{\cal O}(\delta^2) ,
\end{equation}
\begin{equation}
F(r_{j})=F(r_{0})+\epsilon_{j}\delta F^{'}(r_{0})+{\cal
O}(\delta^2) ,
\end{equation}
\begin{equation}
{\cal R}(r_{2})={\cal R}(r_{0})+\delta {\cal R'} (r_{0})+{\cal
O}(\delta^2) ,
\end{equation}
where $\epsilon_{1}=-1$ and $ \epsilon_{2}=+1 $. Inserting Eqs.
$(20-23)$ in the expressions $(21-22)$ and keeping only terms up
to order $ \delta $, we find
\begin{equation}
K^{\chi -}_{\chi}\Bigl|_{\Sigma_{1}}=\frac{1}{R_{0}}
(1+\dot{R}_{0}^2)^{\frac{1}{2}}\left(1+\delta\Bigl(\frac{R^{'}_{0}}
{R_{0}}-\frac{\dot{R}_{0}\dot{R}^{'}_{0}}{1+\dot{R}_{0}^2}\Bigr)\right),
\end{equation}
\begin{eqnarray*}
K^{\chi +}_{\chi}\Bigl|_{\Sigma_{2}}=\frac{1}{R_{0}}
\left(1+\dot{R}_{0}^2-\frac{{\cal R}^2(r_{0})}{R_{0}^2}
\right)^{\frac{1}{2}}
\left(1-\delta\Bigl(\frac{R^{'}_{0}}{R_{0}}-\frac{\dot{R}_{0}\dot{R}^{'}_{0}
-\frac {{\cal R'}(r_0){\cal
R}(r_0)}{R_{0}^2}+\frac{R^{'}_{0}{\cal R}^2 (r_{0})}{R_{0}^3}}
{1+\dot{R}_{0}^2-\frac{{\cal R}^2(r_{o})}{R_{0}^2}}\Bigr)\right),
\end{eqnarray*}
\begin{equation}
\end{equation}
\begin{eqnarray}
 K^{\chi +}_{\chi}\Bigl|_{\Sigma_{1}}=\frac{1}{R_{0}}
\left(1+\dot{R}_{0}^2-\frac{F(r_{0})}{R_{0}^2}
\right)^{\frac{1}{2}}
\left(1+\delta\Bigl(\frac{R^{'}_{0}}{R_{0}}-\frac{\dot{R}_{0}\dot{R}^{'}_{0}
-\frac{F^{'}(r_0)}{2R_{0}^2}+\frac{R^{'}_{0}F(r_{0})}{R_{0}^3}}
{1+\dot{R}_{0}^2-\frac{F(r_{o})}{R_{0}^2}}\Bigr)\right),
\end{eqnarray}
\begin{eqnarray}
 K^{\chi-}_{\chi}\Bigl|_{\Sigma_{2}}=\frac{1}{R_{0}}
\left(1+\dot{R}_{0}^2-\frac{F(r_{0})}{R_{0}^2} \right)^\frac{1}{2}
\left(1-\delta\Bigl(\frac{R^{'}_{0}}{R_{0}}-\frac{\dot{R}_{0}\dot{R}^{'}_{0}
-\frac{F^{'}(r_0)}{2R_{0}^2}+\frac{R^{'}_{0}F(r_{0})}{R_{0}^3}}
{1+\dot{R}_{0}^2-\frac{F(r_{o})}{R_{0}^2}}\Bigr)\right),
\end{eqnarray}
where $ R_{0}=R(r_{0},\tau) $.  Substituting Eqs. $(26-29)$ back
into Eq. $(17)$ and noting that for the metric LTB, $ F(r_{0}) $
is just the square of the Schwarzschild radius of the part of the
spherical thick brane within the comoving surface $ r_{0} $
denoted by $ {\cal R} (r_{0}) $, we obtain after some
simplification and rearrangement the equation of motion of the
dust thick brane written up to the first-order in $\delta $
\begin{eqnarray}
\alpha-\beta= \delta \frac{F^{'}(r_{0})}
{R_{0}^2\left(1+\dot{R}_{0}^2-\frac{F(r_{0})}{R_{0}^2}\right)^{\frac{1}{2}}}\quad\quad\quad\quad\quad\quad
\end{eqnarray}
\begin{eqnarray*}
\hspace*{3.5cm}-\delta \left( \frac{R^{'}_{0}}{R_{0}}(\alpha -
\beta)+\dot{R}_{0}\dot{R}^{'}_{0} (\frac{\alpha -
\beta}{\alpha\beta})+\frac{{\cal R}(r_{0})}{\beta
R_{0}^2}\Bigl({\cal R}^{'}(r_{0})+\frac{R^{'}_{0}}{R_{0}}{\cal
R}(r_{0})\Bigr)\right),
\end{eqnarray*}
with
\begin{equation}
\label{alpabeta2} \alpha\equiv(1+\dot{R}_{0}^2)^\frac{1}{2}
,\quad\quad  \beta\equiv\left(1+\dot{R}_{0}^2 -\frac{{\cal R}^2
(r_{0})}{R_{0}^2}\right)^\frac{1}{2}.
\end{equation}
Now, we would like to explore the zero-thickness limit of our
formalism applied to the collapse of a dust thick brane in
vacuum. In order to do this, we take the following definition for
the four dimensional energy density of the infinitely thin brane
[9]
\begin{equation}
\rho=\int_{-\epsilon}^\epsilon\check{\rho}(r,\tau)dn ,
\end{equation}
where $ n $ is the proper distance in the direction of the normal
$ n_{\mu} $ and $ 2\epsilon$ is the proper thickness of the shell,
with the metric $(1)$, the definition $(32)$ takes the form
\begin{equation}
\rho=\int_{-\delta}^\delta
\check{\rho}(r,\tau)\frac{R^{'}(r,\tau)}{\left(1+E(r)\right)^{\frac{1}{2}}}dr.
\end{equation}
Using Eqs. $(9-10)$, we see that Eq. $(33)$ can therefore be
written as
\begin{equation}
16\pi G_{5}
\rho=\int_{-\delta}^\delta\frac{3F^{'}(r)}{R^3\left(1+\dot{R}^2-\frac{F(r)}{R^2}
\right)^{\frac{1}{2}}}dr.
\end{equation}
Now, by using Eqs. $(23-24)$, we may integrate Eq. $(34)$ up to
the first order in $ \delta $ to yield
\begin{equation}
16\pi G_{5}\rho =
\frac{3F^{'}(r_{0})}{R^3_{0}\left(1+\dot{R}^2_{0}
-\frac{F(r_{0})}{R_{0}^2} \right)^\frac{1}{2}}2\delta + {\cal
O}(\delta^2).
\end{equation}
Substituting $(35)$ back into Eq. $(30)$, we find the following
result
\begin{equation}
\label{alpabeta3}
 \alpha - \beta =
 \frac{8\pi G_{5}}{3}\rho
R_{0}-\delta \left(\frac{R^{'}_{0}}{R_{0}}(\alpha -
\beta)+\dot{{R}_{0}} \dot{{R}^{'}_{0}}(\frac{\alpha - \beta}
{\alpha\beta})+\frac{{\cal R}(r_{0})}{\beta R_{0}^2}\Bigl({\cal
R'} (r_{0})+\frac{R^{'}_{0}}{R_{0}}{\cal R}(r_{0})\Bigr)\right).
\end{equation}
Note that in the zero thickness limit of the brane as $ \delta
\rightarrow 0$, the second term on the right hand side of Eq.
$(36)$ is regular and goes to zero such that Eq. $(36)$ reduces to
the Israel's equation of motion written for the dust thin brane in
vacuum(see [9,10] for the dust thin shell in 4D gravity). To see
the explicit effect of the finite thickness on the thick brane
dynamical equation $(36)$ obtained up to the first order in $
\delta $, we rewrite it as
\begin{equation}
\label{alpabeta4} \alpha-\beta=\frac{8\pi G_{5}}{3}\varrho R_{0},
\end{equation}
where $ \varrho $ is defined by
\begin{equation}
\varrho=\rho-\frac{3\delta}{8\pi
G_{5}R_{0}}\left(\frac{R^{'}_{0}}{R_{0}}
(\alpha-\beta)+\dot{R}_{0}\dot{R}^{'}_{0}(\frac{\alpha-\beta}{\alpha\beta})
+\frac{{\cal R}(r_{0})}{\beta R_{0}^2}\Bigl({\cal R'}
(r_{0})+\frac{R^{'}_{0}}{R_{0}}{\cal R}(r_{0})\Bigr)\right).
\end{equation}
Equation (37) has the same form as the well-known Israel's
junction equation for a thin brane with the effective energy
density $\varrho$.\\
 Now, for a dust brane starting its collapse at rest, the
velocity $\dot{R}_{0}$, being negative during the collapse,
becomes more negative with increasing $r$, i.e. $\dot{R'}_{0}< 0$.
Therefore, the combination $\dot{R}_{0}\dot{R'}_{0}$ is positive.
On the other hand, the Schwarzschild radius of the brane layers
are increased with $r$ so that ${\cal R'} (r_{0})>0$. We therefore
see that all terms within the bracket on the right hand side of
Eq. (38) are positive, meaning that $\varrho<\rho$.\\
 Now, solving Eq. (37) for $\dot{R}^{2}$ we get
\begin{equation}
\dot{R}_{0}^{2} =\frac{16}{9}\pi^{2} G_{5}^{2} R_{0}^{2}
\varrho^{2}+\frac{9{\cal R}^{4}
(r_{0})}{256\pi^{2}G_{5}^{2}R_{0}^{6} \varrho^{2}} +\frac{{\cal
R}^2 (r_{0})}{2R_{0}^2} -1.
\end{equation}
From Eq. (39), it follows that for a given $R_{0}$ and ${\cal R}
(r_{0})$, as long as $R_{0}>{\cal R}(r_{0})$, smaller $\varrho
(\varrho <\rho )$ leads to a larger $\dot{R}^{2}$. Therefore, the
first-order thickness corrections to the Israel thin brane
approximation yield a faster collapse of the dust brane in vacuum.
\subsection*{III. Conclusions}
\hspace*{0.5cm}Motivated by the idea that our universe  is a slice
of a higher dimensional spacetime, we have applied our earlier
approach based on the Darmois matching conditions to obtain the
dynamical equation describing the motion of a spherical thick
3-brane for which the matter on it is in the form of a dust fluid,
the bulk is empty, and no cosmological constants were
introduced.\\
It turns out that the effect of thickness up to the first order in
the brane thickness is to speed up the collapse of the brane in
vacuum. Our results have a well defined zero thickness limit which
are just the Israel thin shell approximation for 3-branes. In the
present work, we have made a number of assumptions to simplify the
analysis. It would be interesting to see how the results may be
changed by relaxing some of these assumptions. For example, what
would be changed if instead of comoving thickness of the brane the
proper brane thickness is assumed to be constant in time. What
would be the effect of taking into account the peculiar velocity
of the brane layers with respect to the background?  An obvious
generalization would also be a non-zero cosmological constant in
the bulk. A non-zero pressure fluid in the brane is another
interesting generalization.\\

\end{document}